\documentclass[aps, prl,twocolumn,showpacs,amsmath,amssymb,superscriptaddress, longbibliography, 10pt]{revtex4-1}

\usepackage{tgtermes}
\usepackage[utf8]{inputenc}
\usepackage{bm}
\usepackage{color}
\usepackage{makeidx}
\usepackage{amsfonts}
\usepackage{amssymb}
\usepackage{amsthm}
\usepackage{graphicx}
\usepackage{epstopdf}
\usepackage{subfigure}
\usepackage{ulem}
\usepackage{amsmath,amssymb}
\usepackage[colorlinks=true,citecolor=blue,urlcolor=blue]{hyperref}
\usepackage{color}

\hypersetup
	{ colorlinks,%
		citecolor=blue,%
		linkcolor=blue,%
	}

\newcommand{\bea}{\begin{eqnarray}}
\newcommand{\ea}{\end{eqnarray}}
\newcommand{\eea}{\end{eqnarray}}

\newcommand{\sumint}[1]

\begin{document}

\title{Dynamically observing the spectra of quantum droplets in optical lattice}
\author{Yuhang Nie}
\affiliation{Shaanxi Key Laboratory for Theoretical Physics Frontiers, Institute of Modern Physics, Northwest University, Xi'an, 710127, China}
\author{Jun-Hui Zheng}
\email{junhui.zheng@nwu.edu.cn}
\affiliation{Shaanxi Key Laboratory for Theoretical Physics Frontiers, Institute of Modern Physics, Northwest University, Xi'an, 710127, China}
\affiliation{School of Physics, Northwest University, Xi'an, 710127, China}
\affiliation{Peng Huanwu Center for Fundamental Theory, Xi'an 710127, China}
\author{Tao Yang}
\email{yangt@nwu.edu.cn}
\affiliation{Shaanxi Key Laboratory for Theoretical Physics Frontiers, Institute of Modern Physics, Northwest University, Xi'an, 710127, China}
\affiliation{School of Physics, Northwest University, Xi'an, 710127, China}
\affiliation{Peng Huanwu Center for Fundamental Theory, Xi'an 710127, China}

\date{\today}

\begin{abstract}

Optical lattice plays an important role on stability and dynamics of quantum droplets.  In this letter, we investigate the Bogoliubov excitation spectrum of quantum droplets in optical lattice in the thermodynamic limit.  We classify the collective excitations as synchronous modes, Bloch phononic modes, and site-density imbalanced modes. For synchronous modes, we measure the dipole oscillation frequencies by quench dynamics with a sudden shift of the optical lattice, and the breathing frequencies by Floquet dynamics with a periodic change of the lattice depth. Bloch phononic modes are observable from the Landau critical velocity of the droplets. We further discuss the instability induced by the site-dependent density fluctuations, and calculate the critical filling of atoms where the growth of lattice vacancy breaks down the translational symmetry of the system. This work makes essential steps towards measuring the excitation spectrum and understanding the superfluid nature of quantum droplets in optical lattice.
\end{abstract}


\maketitle

Quantum droplets are self-bound states formed through a balance of particle-particle interactions without any external trapping, where attractions bring the constituent particles together and repulsions stabilize the droplets from collapse\,\cite{Bender2003,Dalfovo2001,Bulgac2002,Petrov2015,Petrov2016,Baillie2016,Chomaz2016,Staudinger2018,Sekino2018,Wachtler2016,Zin2018,Ilg2018,Luo2021,Li2017}. In experiments with ultracold atoms, quantum droplets have been achieved in single-component dipolar Bose-Einstein condensates (BECs) and binary BECs with interspecies attractions \cite{Schmitt2016,Cabrera2017,Cheiney2018,Ferrier2016,Kadau2016,Semeghini2018,Ferrier2018}, where the competition arises from the mean-field interaction and the Lee-Huang-Yang correction of the quantum zero-point energy \cite{Petrov2015,Petrov2016,Baillie2016,Chomaz2016,Wachtler2016}. The roles of these interactions in the formation of quantum droplets depend strongly on the dimensionality of the system\,\cite{Petrov2016}. It was shown that with decreasing number of atoms $N$, free droplets become metastable and eventually disappear when $N$ is smaller than a critical value. The excitations of free particle emissions, discrete monopole (breathing) modes, and surface ripplon modes of three-dimensional (3D) free droplets have also been well studied\,\cite{Petrov2015}.

In systems with external traps, quantum droplets are stabilized further by suppression of emission of free particles, and the trapping potential helps to manipulate the droplet mechanically \cite{Dong2021,Tengstrand2019,Hu2020}. The effect of a weak harmonic trapping potential on surface modes of 3D quantum droplets was studied in\,\cite{Hu2020}. However, the trapping confinements somehow obscure the characteristics of quantum droplets and become an obstacle to observe the formation and inflation of droplets. Optical lattice is an effective mean to stabilize droplets and their topography, which integrates not only the advantages of a trap but also the capability to show the growth of quantum droplets\,\cite{Zheng2021}. To understand the features of quantum droplets, accurate frequency measurement of collective modes is a powerful tool. For example, the collective mode measured in arrays of dipolar droplets clearly shows the symmetry breaking and the supersolid nature of the system\,\cite{Tanzi2019,Guo2019}. Therefore, it is interesting to study the collective excitations of quantum droplets in optical lattices, which would be an excellent way to characterize this fascinating matter. It is also helpful to probe the interplay between light and condensed matter, which usually brings novel phenomena\,\cite{Morsch2006,Chen2016,Zaera2018,Zhou2019,Dong2020,Chen2021,Zhao2021,Zhang2019}.

In this letter, we consider an infinitely large system of two-dimensional (2D) quantum droplets in an optical lattice. Therefore, the boundary effects of a cluster of lattice quantum droplets existing in a finite-size system are negligible, and one can focus on the universal behaviors of droplets. The ground state and the Bogoliubov excitation spectrum of the quantum droplets for different lattice depths are calculated by using the imaginary time evolution method and the exact diagonalization method, respectively. Bloch phononic modes are observable from the Landau critical velocity of the droplets. The properties of quench dynamics and Floquet dynamics of the droplets are investigated by suddenly shifting the lattice potential and periodically driving the system. The oscillation frequency and intrinsic breathing frequency obtained agree very well with the Bogoliubov excitation spectrum and are measurable in ongoing experiments. Moreover, number imbalances among different sites of the lattice may induce instability of the droplets. We find that below a critical filling, vacancies in the lattice appear spontaneously, and the translational symmetry of the system breaks down, which is the prelude of producing self-assembly of quantum droplets from different sites.

Considering a binary BEC loaded in a 2D optical lattice and taking into account the Lee-Huang-Yang correction, the two components of the BEC with the same scattering length and particle number, can be described by a common wavefunction. This wavefunction, in a rescaled form, follows the extended Gross-Pitaevskii equation (GPE) in the dimensionless form\,\cite{Li2018,Zheng2021},
\begin{equation}\label{2}
i{\partial_t \phi}=-\frac{1}{2} \nabla^{2} \phi+V(x, y) \phi+|\phi|^{2} \phi \ln |\phi|^{2},
\end{equation}
where the 2D optical lattice potential reads
\begin{equation}\label{3}
V(x, y)=V_{0}\left[\sin ^{2}\left(\frac{\pi}{d} x + \delta\right)+\sin ^{2}\left(\frac{\pi}{d} y\right)\right].
\end{equation}
Here, $V_0$ is the optical lattice depth, $d$ is the lattice constant, and $\delta$ is used to shift the entire lattice. Without loss of generality, we will set $d=5$ in the following calculations. 

We focus on the thermodynamic limit with an infinite size of the system and assume that all sites are occupied by droplets, so the system initially preserves translational symmetry. In this case, the particle number per lattice site,
\begin{equation}\label{6}
N_s=\iint_{d^2}|\phi|^2 {d} x {d} y,
\end{equation}
refers to the filling of the system. Meanwhile, the energy per site is
\begin{equation}\label{5}
E_s=\frac{1}{2} \iint_{d^2}\left[|\nabla \phi|^2+2V|\phi|^2+|\phi|^4 \ln \left(\frac{|\phi|^2}{\sqrt{e}}\right)\right] {d} x {d} y.
\end{equation}


\begin{figure}[tpb]
  \centering
  \includegraphics[width=0.97\columnwidth]{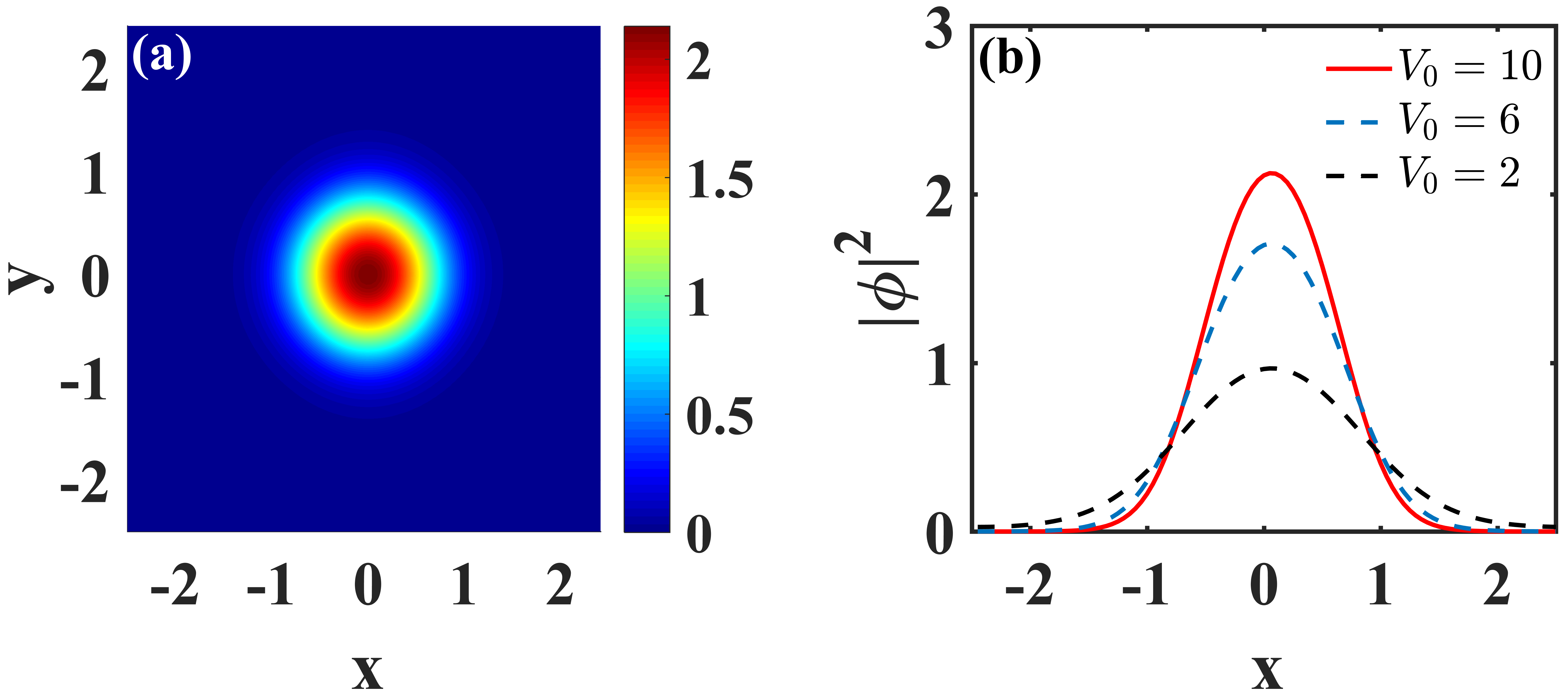}
  \caption{Density profiles of quantum droplets in one cell of an optical lattice with filling $N_s=3.5$. (a) The top view of the density distribution for $V_0=10$. (b) The cross sections of the density distribution at $y=0$ for $V_0 = 2, 6, 10$, respectively.}\label{fig1}
\end{figure}

In the absence of the external potential $V$, the density distribution of a quantum droplet is flat everywhere except at boundaries for a large total particle number $N$. After neglecting the surface energy, minimizing the total energy $E\simeq (N/2) \rho_0 \ln (\rho_0 /\sqrt{e})$ gives the favored density $\rho_0 = \exp(-1/2)\simeq 0.61$\,\cite{Petrov2016}. In an optical lattice, the ground-state wavefunction $\phi_0$ can be obtained by numerically solving Eq.\,\eqref{2} in the imaginary time evolution. 
For wavefunctions $\zeta\propto\xi \doteq -\frac{1}{2} \nabla^{2} \phi+V(x, y) \phi+|\phi|^{2} \phi \ln |\phi|^{2}$ with $\zeta$ being normalized to $N_s$, we obtain the final self-consistent ground-state wavefunction with good precision $\langle \zeta | \phi \rangle/N_s \geq 1-10^{-6}$. The chemical potential $\mu$ is determined by $\mu = \sqrt{\langle \xi | \xi \rangle/N_s}$. We note that high enough accuracy will ensure the correction of the Bogoliubov spectrum of the system.

In Fig.\,\ref{fig1}(a), we show the density profile of the ground state quantum droplet in a single unit cell for $V_0=10$ and $N_{s}=3.5$. The droplets at different sites are the same in size and shape. The existence of the optical lattice significantly changes the flat-top structure of the density profile. With increasing depth of the lattice, the peak density increases accordingly, while the radius of the droplet decreases, as shown in Fig.\,\ref{fig1}(b). The peak densities of the droplets in the lattice are all higher than $\rho_0$ to minimize the energy of the system. With decreasing $V_0$, the system may undergo a transition from effectively repulsive interactions to attractive interactions when the peak density becomes smaller than 1. For cases with large $V_0$, the central part of the droplet is dominated by the repulsive interaction, while the attractive interaction dominates the rim of the droplet.

The collective excitation modes can be obtained by considering a small fluctuation around the ground-state wavefunction $\phi_0(x, y, t)=\psi_0(x, y) e^{-i \mu t}$, i.e., $\phi = (\psi_0 + \psi_1)e^{-i \mu t}$. By expanding Eq.\,\eqref{2} with respect to $\psi_1$ and $\psi_1^*$, we obtain the evolution equation of the fluctuation $\psi_1$ under the linear approximation,
\begin{eqnarray}\label{eq7}
  i{\partial_t \psi_1}&=& -\frac{1}{2} \nabla^{2} \psi_1 +V\psi_1+(2 \psi_0^2 \psi_1 + \psi_0^2 \psi^*_1)  \ln \psi_0^{2} \notag\\
  &&+ \psi_0^2 (\psi_1 + \psi^*_1)-\mu \psi_1,
\end{eqnarray}
and its conjugate equation.
Now, we consider the excitation mode $\psi_1 = \tilde u e^{-i\omega t}+ \tilde v^* e^{i\omega t}$, and assume that the wavefunction takes the form of the Bloch wave, i.e., $ \tilde u(\bm r) = e^{i\bm k\cdot \bm r}  u_{\bm k}(\bm r)$ and $ \tilde v(\bm r) = e^{i\bm k\cdot \bm r}  v_{\bm k}(\bm r)$, where both $u_{\bm k}(\bm r)$ and $u_{\bm k}(\bm r)$ are periodic in space, similar to the ground state and the lattice potential. As a result, we obtain
\begin{equation}\label{eq8}
  \omega_{\bm k} \begin{pmatrix}
   u_{\bm k}\\
   v_{\bm k}
  \end{pmatrix}
  =
    \begin{pmatrix}
  A &  B\\
 - B & - A
  \end{pmatrix}
     \begin{pmatrix}
   u_{\bm k}\\
  v_{\bm k}
  \end{pmatrix},
\end{equation}
where
\begin{eqnarray}\label{x}
  A &=& \frac{1}{2} \bm k^2 - i \bm k \cdot \nabla  -\frac{1}{2} \nabla^{2} +V +2 \psi_0^2   \ln \psi_0^{2} + \psi_0^2-\mu, \\
  B &=& \psi_0^2 \ln \psi_0^{2} + \psi_0^2.
\end{eqnarray}

\begin{figure}
  \centering
  \includegraphics[width=\columnwidth]{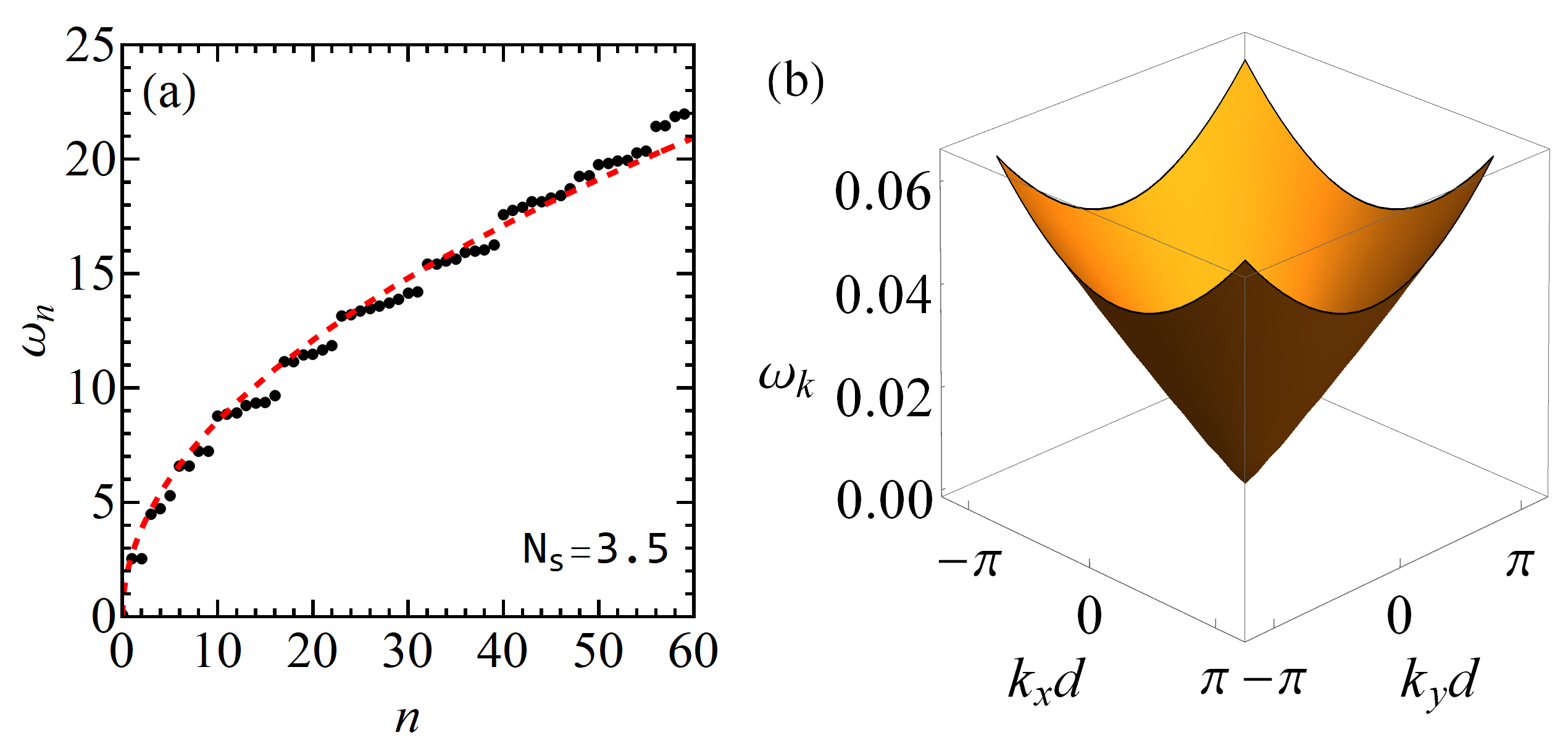}\\
  \caption{Bogoliubov spectrum of collective excitations of lattice quantum droplets for $V_0=10$. (a) The synchronous excitation modes.
  The dashed red line is the fitting function $ n = \alpha\omega^2$ with a parameter $\alpha$. (b) The Bloch phononic spectra. }\label{fig2}
\end{figure}

With the obtained $\psi_0$ and the chemical potential $\mu$, we numerically calculate the Bogoliubov spectrum of the collective excitations using the exact diagonalization method. In Fig.\,\ref{fig2}(a), we show the excitation spectrum for $\bm k =0$. In this case, the wavefunctions of the excited states possess translational invariance, which can be named as synchronous excitation modes of the quantum droplets in the lattice. We can find that the total number of states below energy $\omega$ approximately satisfies $n \sim \alpha {\omega}^2$ with $\alpha$ being a fitting parameter. This result is similar to that of the number of states of a noninteracting system in a harmonic trap with frequency $\omega_h$. The latter has spectra $\omega = (2 n_r +|l|)\omega_h$, where $n_r$ is the radial quantum number and $l$ is the angular momentum quantum number. We note that this fitting is still valid for the excitation spectrum up to $n=60$, and for excitations with higher energy or a shallow lattice, the $\omega_n$-$n$ relation deviates significantly from the fitting function.

From the energy spectrum Fig.\,\ref{fig2}(a), we can clearly see that the first excitation is double degenerate, which is due to the $C_4$ symmetry in the $x$-$y$ plane. These two excitations correspond to the excitation modes $(n_r, l) = (0, \pm 1)$ of a harmonically trapped noninteracting system. The next three excitations correspond to the degenerate excitation modes $(n_r, l) = (1,0), (0, \pm 2)$. With repulsive interaction, the spectrum becomes $\omega = \omega_h \sqrt{2 n_r +2 n_r^2 +2 n_r |l| + |l|}$ where the breathing mode $(1,0)$ and the other two are not degenerate \cite{Zheng2015}. In addition, the $U(1)$ rotational symmetry is broken on a lattice, which will lead to further splitting of the rest two modes. In Fig.\,\ref{fig2}(b), {we plot the phononic spectra of the Bloch waves in the first Brillouin zone. In experiments, it is convenient to determine phononic excitation spectra by measuring the Landau critical velocity of the superfluid. This velocity ($\simeq 0.073$) is the gradient of the dispersion relation along different directions at $\bm{k}=0$.} In contrast to the synchronous excitation modes, the Bloch waves focus on the phase fluctuation of the system --- the density distribution in each lattice remains the same but the phases vary. We also find that the group velocities of phononic excitations along different directions are almost the same. All the spectra shown in Figs.\,\ref{fig2}(a) and \ref{fig2}(b) are real and positive, indicating that the system is stable under synchronous excitations or phase fluctuations.

\begin{figure}[t]
  \centering
  \includegraphics[width=\columnwidth]{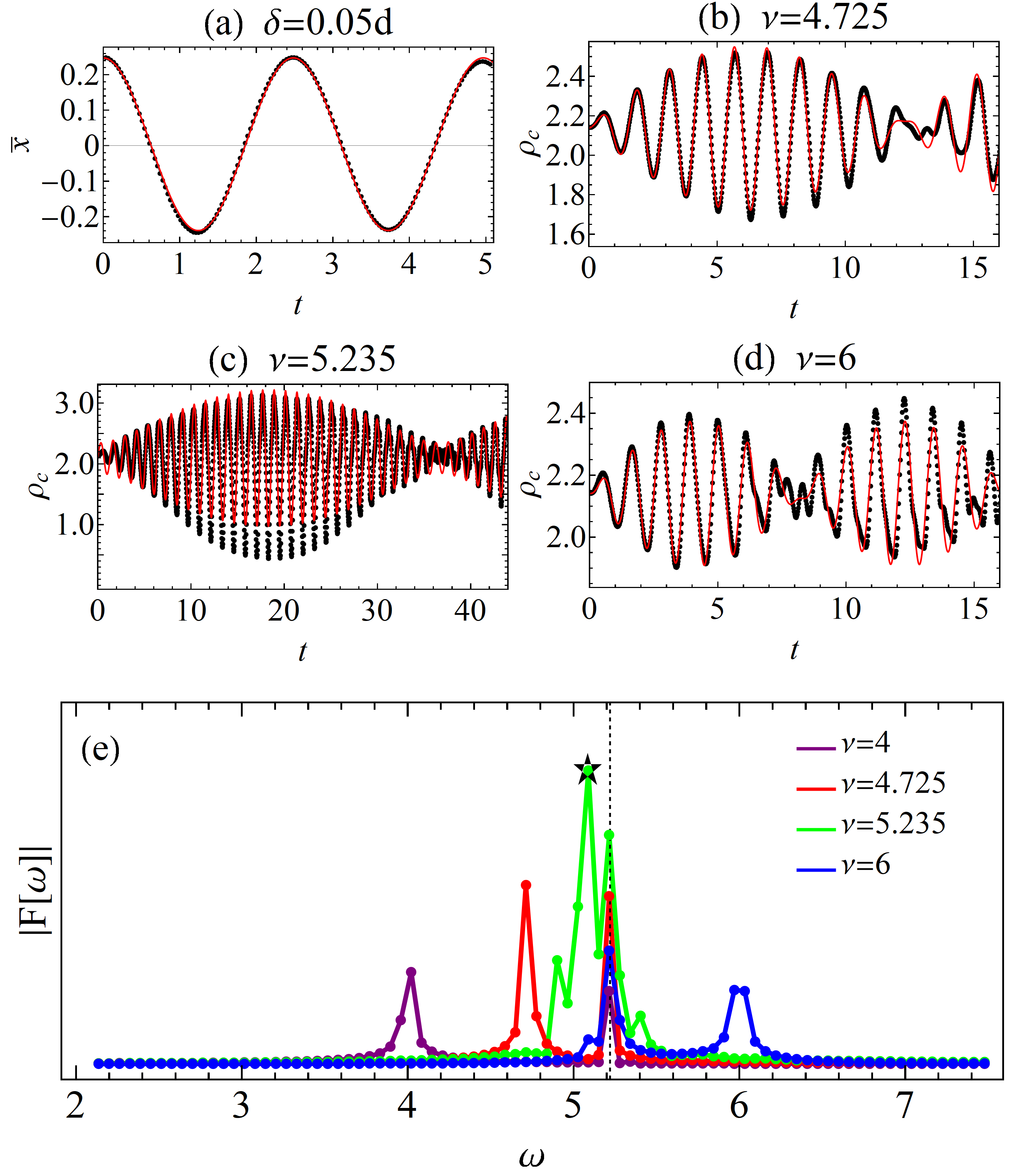}
  \caption{(a) The center-of-mass of the droplet changes over time during quench dynamics when the lattice is suddenly shifted. The red solid line is the fitting result with the cosine function $\cos(\Omega_1 t)$. (b-d) The evolution of the density ($\rho_c$) at the center of each site for different driving frequencies $\nu$. The red solid lines show the fitting results of $a_0 + a_1 \sin (\nu t) + a_2 \sin (\omega t)$, where $\omega$ equals the breathing frequency ($\omega_{n}=5.235$) for (b) and (d), but is shifted to $\tilde\omega \simeq 5.06$ for the resonant case (c). (e) The spectrum function obtained from the Fourier transformation of $\rho_c(t)$ over the interval $t \in [0,100]$.}\label{fig3}
\end{figure}

The synchronous excitation spectrum can be obtained from the dynamics of the droplets. In the following, we will study the quench dynamics and Floquet dynamics of the droplet in the lattice. By suddenly shifting the lattice for 5$\%$ of the lattice constant along the $x$- ($y$-) direction, we identify the oscillation mode of the droplets under quench dynamics. As a result, the center-of-mass of the droplets deviates from the minimum of each cell of the lattice potential. The droplets start to oscillate in the $x$- ($y$-) direction. The typical oscillating behavior of the center-of-mass ($\bar x$) of the droplet is $\cos(\Omega_1 t)$ as shown in Fig.\,\ref{fig3}(a). The magnitude of $\Omega_1$ can be obtained by fitting the evolution of the oscillation.

By periodically modulating the potential depth with
\begin{equation}
  V_0 \rightarrow V_0+\delta V \sin (\nu t),
\end{equation}
we can identify the breathing mode of the droplets under Floquet dynamics. The density of particles $\rho_c$ at the center of each cell of the lattice varies with the potential depth, which can be employed to characterize the breathing behavior. For different initial potential depths $V_0$, we set the modulation amplitude to be $\delta V = 5\%V_0$, and tune the driving frequency $\nu$. For $V_0 = 10$ as shown in Figs.\,\ref{fig3}(b)-\ref{fig3}(d), the time evolutions of $\rho_c$ exhibit regular oscillating behaviors with different amplitudes and periods by applying three typical driving frequencies. When the amplitude is sufficiently large, it becomes asymmetric (see Fig.\,\ref{fig3}(c)) due to nonlinear effects.

\begin{figure}[tbp]
  \centering
  \includegraphics[width=0.95\columnwidth]{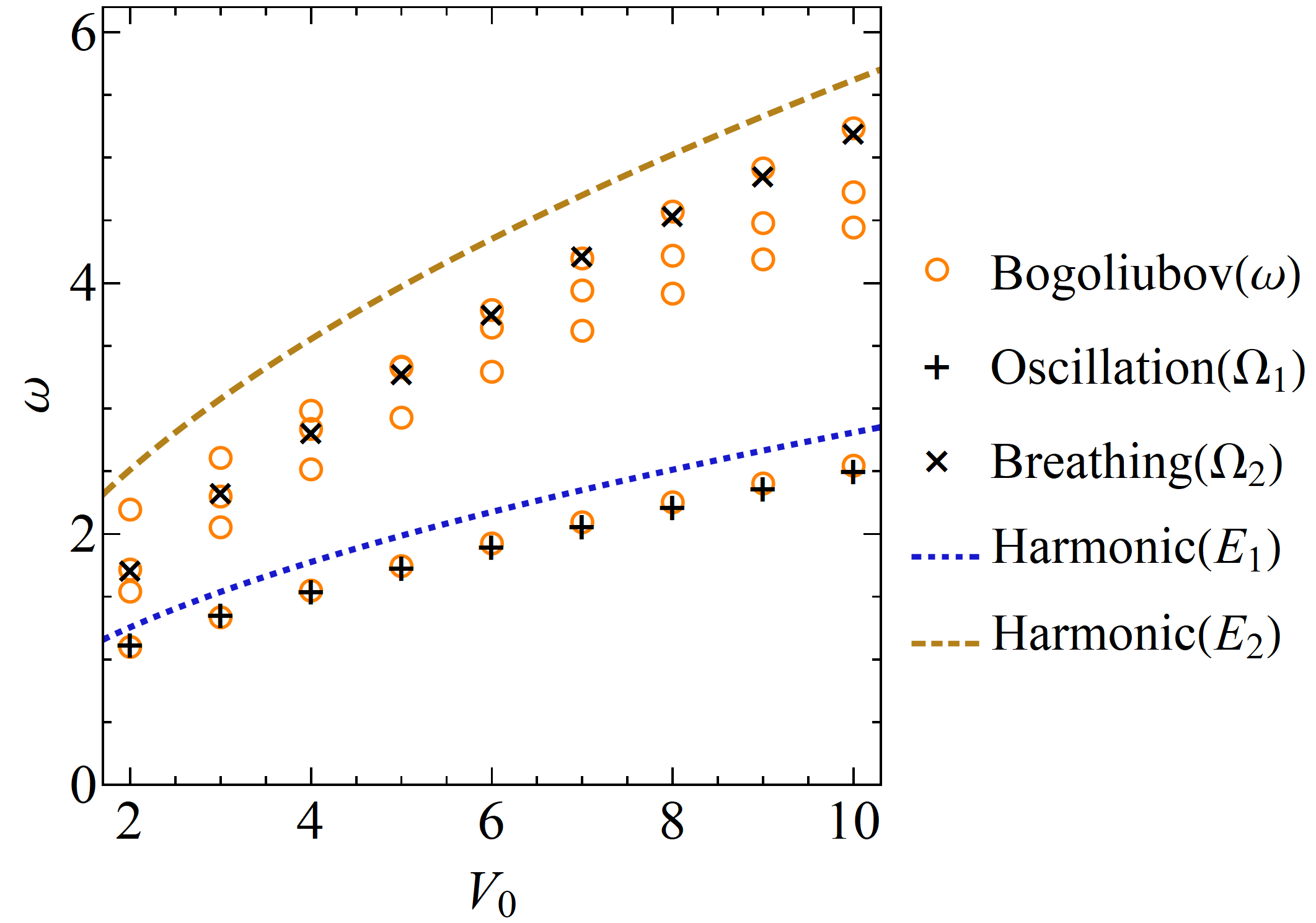}
  \caption{The Bogoliubov spectra (orange circles), the dynamical dipole oscillation frequencies (black `$+$'), the intrinsic breathing frequencies (black `$\times$'), and the excitation spectra  (dashed lines, $E_2 = 2E_1 = 2\omega_h$) in the harmonic trap with frequency $\omega_h = \sqrt{2V_0}\pi/d$. }\label{fig4}
\end{figure}

We further apply the Fourier transform to obtain the spectrum functions of the above oscillations with $t\in [0,100]$. As shown in Fig.\,\ref{fig3}(e), we observe that when the driven frequency is away from the spectrum of the `second' synchronous excitation (the `first' synchronous excitation is the oscillation excitation), there are two frequencies that dominate the spectrum function. One frequency is nothing but the driven frequency itself, and the other one remains unchanged when tuning $\nu$. {Therefore, we match this intrinsic frequency $\Omega_2$ ($\simeq 5.22$, see the black dashed line) to the frequency of the breathing mode ($\omega_{n}=5.235$) of the droplets. Basically, the closer the driving frequency is to $\Omega_2$, the greater the oscillation amplitude, and the longer the period (see Figs.\,\ref{fig3}(b)-\ref{fig3}(d)). The fitting function $a_0 + a_1 \sin (\nu t) + a_2 \sin (\omega_{n} t)$ agrees very well with the real-time evolution $\rho_c(t)$, as shown in Figs.\,\ref{fig3}(b) and \ref{fig3}(d). However, in the near-resonant regime ($\nu -\Omega_2 \ll \Omega_2$), the spectrum function indicated by the green solid line ($\nu=5.235$) in Fig.\,\ref{fig3}(e) gives more structures around $\Omega_2$, and the obtained frequency ($\simeq 5.086$, marked by a black star) besides the driven frequency, deviates from the intrinsic frequency slightly. Fitting $\rho_c(t)$ with $a_0 + a_1 \sin (\nu t) + a_2 \sin (\tilde\omega t)$ gives $\tilde\omega \simeq 5.06$ and the resulting function is shown in Fig.\,\ref{fig3}(c). We attribute the emergence of $\tilde\omega$ to the hybridization effect of the frequencies of driving and resonating}.

In Fig.\,\ref{fig4}, {we plot the oscillating frequency $\Omega_1$ (marked by black `$+$'), the intrinsic frequency $\Omega_2$ (marked by black `$\times$'), and the Bogoliubov spectrum of synchronous excitations (marked by orange circles) with respect to the lattice depth $V_0$}. We can see that $\Omega_1$ fits the Bogoliubov spectrum very well, and $\Omega_2$ is located in one of the three `second' synchronous excitation modes which corresponds to the breathing mode and is about twice that of $\Omega_1$ for $V_0\geq 5$. The dashed solid lines refer to the oscillating excitation in a harmonic trap with frequency $\omega_h = \sqrt{2V_0}\pi/d$. This harmonic trap is the second-order approximation of the lattice potential, whose first excitation (dipole oscillation) is $E_1 = \omega_h$ with two degenerate modes, while the breathing excitation is $E_2 = 2 \omega_h$ {for cases with repulsive interaction or without interaction. The Bogoliubov spectrum is generally smaller than that in the harmonic trap, as the confinement effect from the lattice potential is weaker.}

\begin{figure}[t]
  \centering
  \includegraphics[width=\columnwidth]{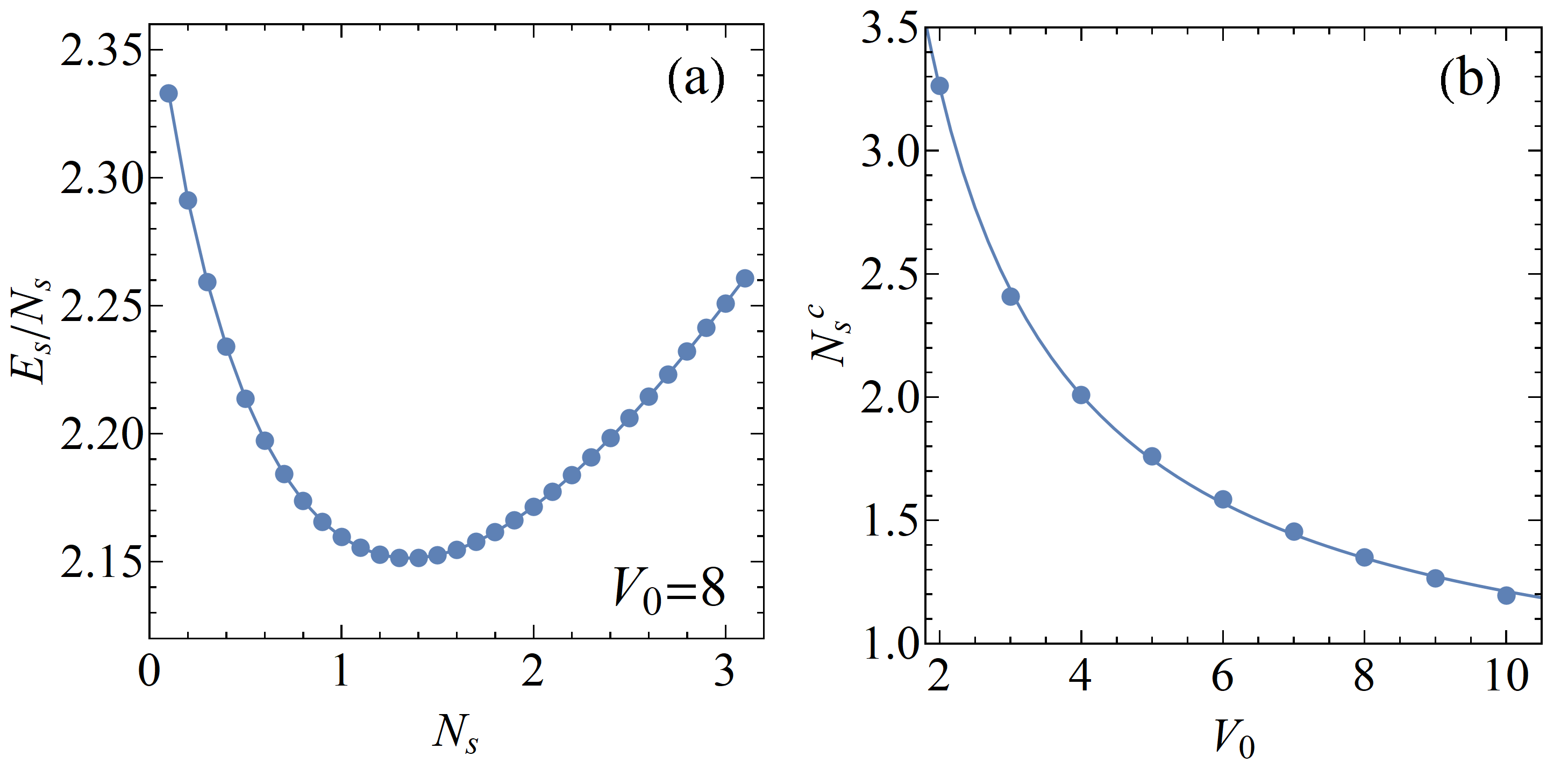}
  \caption{ (a) The energy per atom $E_s/N_s$ varying with the filling $N_s$. (b) The critical filling as a function of lattice depth $V_0$. The solid line is the fitting result of $b_0 + b_1/(V_0 + b_2)$ with $b_0 = 0.661$, $b_1=5.59$, and $b_2=0.156$. }\label{fig5}
\end{figure}

In our calculations, we have assumed that the density distribution of the droplet at each lattice site remains the same. Therefore, the density imbalance between different lattices is completely ignored. For cases with a large filling, the repulsive interaction dominates and the density fluctuations are strongly suppressed. The validity of the density balance assumption is guaranteed. However, for cases with low filling, the attractive interaction becomes stronger. In this situation, small density fluctuations may induce the emergence of self-binding, {leading to the appearance of vacancies in the lattice or even the formation of clusters of quantum droplets from different sites}.

Note that when vacancies start to appear in some sites of the lattice, the filling of particles in other sites will increase to keep the total number of particles constant. Here, we introduce the single atom energy, $\varepsilon \doteq E_s/N_s$ \cite{Petrov2016}. If $\varepsilon$ decreases with increasing $N_s$, it means that the system with vacancies has lower energy. Therefore, the translational symmetry of the system will break down. Otherwise, the system with even filling in all sites is stable. In Fig.\,\ref{fig5}(a), we plot the single-atom energy $\varepsilon$ as a function of the filling $N_s$ for different lattice depths $V_0$. The figure shows that $\varepsilon$ decreases first and then increases as $N_s$ increases. The turning point is the critical filling for the phase transition. In Fig.\,\ref{fig5}(b), we plot the critical filling as a function of lattice depth $V_0$. {With increasing $V_0$, the critical filling decreases approximately with $b_0 + b_1/(V_0 + b_2)$, where $b_0 = 0.661$, $b_1=5.59$, and $b_2=0.156$}. This is because with the same filling, the density is smaller, and the attractive interaction is stronger in a shallow lattice potential.

In conclusion, optical lattices provide a powerful tool for investigating model systems of quantum droplets in periodic potentials through probing excitation properties and nonlinear dynamics. Here, we have investigated the excitation spectrum, quench and Floquet dynamics, and the stability of quantum droplets in an optical lattice in the thermodynamic limit. We find different excitation modes, including the synchronous excitations of droplets from different lattice sites, the Bloch phononic excitation from the phase fluctuations, and excitations from site-dependent density fluctuations. Both dipole oscillations and breathing modes belong to synchronous excitations. By directly fitting the center-of-mass and the density variation of the droplets, or by spectrum analysis through Fourier transformation, we show that the dynamics faithfully record the frequencies of dipolar oscillations and breathing modes. This provides reliable means to detect the excitation spectra in experiments. {Besides, the measurement of the Landau critical velocity of the superfluid can be used to determine the group velocity of the Bloch phonons.} Moreover, when the site-dependent density fluctuations are taken into consideration, the system with low fillings may become unstable, resulting in the appearance of vacancies in some sites of the optical lattice. This process breaks the translational symmetry and is the precursor to the formation of clusters of quantum droplets. It shows the possibility of simulating the formation of clustered galaxies and cosmic webs using lattice quantum droplets. Meanwhile, with the possible emergence of an array of vacancies, our system is a very good candidate for probing a new supersolid phase \cite{Leggett1970,Chester1970}.

\begin{acknowledgments}
This work is supported by the National Natural Science Foundation of China under grants Nos. 12175180, 11934015 and 12247103, the Major Basic Research Program of Natural Science of Shaanxi Province under grants Nos. 2017KCT-12 and 2017ZDJC-32, and Shaanxi Fundamental Science Research Project for Mathematics and Physics under grant Nos. 22JSZ005 and 22JSQ041. This research is also supported by The Double First-class University Construction Project of Northwest University.
\end{acknowledgments}


\end{document}